\documentclass[prl,a4paper,aps,twocolumn,showpacs,superscriptaddress,groupedaddress]{revtex4}
\usepackage[dvips]{graphicx}
\usepackage{ae}
\usepackage[T1]{fontenc}
\usepackage[ansinew]{inputenc}
\usepackage{amsmath}
\usepackage{amssymb}
\usepackage{graphicx}
\usepackage{color}
\usepackage[colorlinks]{hyperref}
\usepackage{lscape}
\begin{document}
\author{Biswajit Paul}
\email{biswajitpaul4@gmail.com}
\affiliation{Department of Mathematics, South Malda College, Malda, West Bengal, India.}
\title{Nonlocality of three-qubit Greenberger-Horne-Zeilinger-symmetric states}
\author{Kaushiki Mukherjee}
\email{kaushiki_mukherjee@rediffmail.com}
\affiliation{Department of Mathematics, Government Girls' General Degree College, Ekbalpore, Kolkata, India.}
\author{Debasis Sarkar}
\email{dsappmath@caluniv.ac.in}
\affiliation{Department of Applied Mathematics, University of Calcutta, 92, A.P.C. Road, Kolkata-700009, India.}

\begin{abstract}
Mixed states appear naturally in experiment over pure states. So for studying different notions of nonlocality and their relation with entanglement in realistic scenarios, one needs to consider mixed states. In a recent article [Phys. Rev. Lett. \textbf{108}, 020502 (2012)], a complete characterization of entanglement of an entire class of mixed three qubit states with the same symmetry as Greenberger-Horne-Zeilinger state known as GHZ-symmetric states, has been achieved. In this paper we investigate different notions of nonlocality of the same class of states. By finding the analytical expressions of maximum violation value of most efficient Bell inequalities we obtain the conditions of standard nonlocality and genuine nonlocality of this class of states. Also relation between entanglement and nonlocality is discussed for this class of states. Interestingly, genuine entanglement of GHZ-symmetric states is necessary to reveal standard nonlocality. However, it is not sufficient to exploit the same.
\end{abstract}
\date{\today}
\maketitle

\section{I. INTRODUCTION}
Quantum nonlocality is an inherent character of quantum theory where Bell inequalities \cite{Bel} are used as witnesses to test the appearance of the same. Recently analysis of quantum nonlocality has become an interesting topic not only from foundational viewpoint(See \cite{Bru} and references therein) but also has been extensively used in various quantum information processing tasks, quantum communication complexity \cite{Bur}, randomness amplification \cite{Ram}, no-signaling \cite{Bar}, device-independent quantum key distribution \cite{Aci}, device-independent quantum state estimation \cite{Bad,Mck}, randomness extraction \cite{Col,Pir}, etc. There exist several experimental evidences supporting the fact that presence of entanglement is necessary for nonlocality of quantum correlations. But determining which entangled state reveals nonlocality(i.e., violates bell inequality) is a difficult work. Any pure entangled state of two qubits violates Bell-CHSH inequality\cite{Cla,Gsi}, the amount of violation being proportional to the degree of bipartite entanglement\cite{Ppe}. However no such conclusion holds for mixed bipartite entangled states as there exists a class of mixed entangled states which admits a local hidden variable model and this class cannot violate any Bell inequality \cite{Wer,Brr}. Till date nonlocality in two qubit systems has been explored in details. However multiqubit case is much more difficult to analyze. \\
There is an increasing complexity while shifting from bipartite to multipartite systems. This is mainly because of the fact that multipartite entanglement has comparatively much complex and richer structure than that of bipartite entanglement \cite{Guh,Hor}. So any study related to multipartite entanglement or dealing with multipartite nonlocality requires a deeper insight of the physics of many-particle systems which in general differ extensively from that of single or two party systems. However, study of many-particle systems gives rise to new interesting phenomena, such as phase transitions \cite{Zol}
or quantum computing. In this context, it is quite interesting to study the relationship between entanglement and nonlocality for multipartite system. To extend the two qubit relationship between entanglement and quantum  nonlicality, one needs to classify both entanglement and nonlocality in multipartite system. In particular, entanglement of any tripartite state can either be biseparable or genuinely entangled \cite{Guh,Hor}. Nonlocal character of a tripartite system can be categorized broadly in two categories of standard nonlocality and genuine nonlocality \cite{Bru}. In the former case, nonlocality is revealed in atleast one possible grouping of the parties whereas a state is said to be genuinely nonlocal if any possible grouping of parties reveal nonlocality. In \cite{Sli}, \'{S}liwa gave the whole class of Bell inequalities which acts as a necessary and sufficient condition for detecting standard nonlocality. The relation between this notion of nonlocality and tripartite entanglement has been studied for three qubit pure states\cite{Sca,Zuk,Ema,Che,Chi,Syu} where it has been shown that entanglement(biseparable or genuine entanglement) of pure state suffices to produce standard nonlocality. The notion of genuine tripartite nonlocality  has been discussed in \cite{Svt,Gal,Ban}, which represents the strongest form of nonlocality for tripartite systems. There exists relation between genuine tripartite nonlocality and 3-tangle \cite{Cof} (measure of genuine tripartite entanglement) which has been analyzed for some important classes of pure tripartite states \cite{Ban,Gho,Ajo,Ghs,Kau}. Interestingly, Bancal et al. conjectured that all genuinely entangled pure quantum states can produce genuine nonlocal correlations \cite{Ban}. While tripartite nonlocality turns out to be a generic feature of all entangled pure states, the situation becomes much more complex when we consider mixed states as there exists genuine tripartite entangled state which admits a local hidden variable model \cite{Tot,Bow}. In this context, it is interesting to characterize the state parameters for any class of tripartite mixed states on the basis of different notions of tripartite nonlocality and their relation with entanglement. Our paper goes in this direction. Recently, a new type of symmetry for three qubit quantum state was introduced \cite{Elt}, the so called Greenberger-Horne-Zeilinger (GHZ) symmetry. In \cite{Elt}, they provided the whole class of states which has this type of symmetry. This class of states is referred to as GHZ-symmetric states. A complete classification of different types of entanglement of this class of tripartite mixed states is made in \cite{Elt}. In this work we have classified the GHZ-symmetric states on the basis of different notions of tripartite nonlocality so that one can use this class of state in different information theoretic tasks. This helps us to establish the relationship between entanglement and nonlocality for this class of tripartite mixed states. The relation implies that genuine entanglement is necessary to  reveal any type of nonlocality(standard nonlocality or genuine nonlocality) for this class of states.\\

The paper is organized as follows. In Section II, we give a brief introduction to some concepts and results which we will use in later sections. Subsequently, in Section III, we obtain the condition for which GHZ-symmetric states reveal standard nonlocality by deriving the analytical expressions of maximum violation value of the two most efficient facet inequalities. In section IV, we deal with the classification of the class of states on the basis of genuine nonlocality. Section V shows how different types of entanglement are related with different notions of nonlocality for this class of mixed states. Finally we conclude with a summary of our results in Section VI.
\section{II. BACKGROUND}
\subsection{A. GHZ-symmetric three-qubit states}
As an important class of mixed states from quantum theoretical perspective, GHZ-symmetric three-qubit states have been paid much attention \cite{Elt,Sie,Els,Buc}. In particular, in the eight dimensional state space of  three qubit states, the set of GHZ-symmetric states defines a two-dimensional affine section, specifically a triangle of the full eight dimensional set of states \cite{Goy}. In this section, we review the properties of GHZ-symmetric three-qubit states\cite{Elt}. GHZ-symmetric three-qubit states are defined to be invariant under the following transformations: (i) qubit permutations, (ii) simultaneous three-qubit flips (i.e., application of $\sigma_x \otimes \sigma_x \otimes\sigma_x$), (iii) qubit rotation about the z axis of the form $U(\phi_1, \phi_2)= e^{i\phi_1\sigma_z}\otimes e^{i\phi_2\sigma_z}\otimes e^{-i(\phi_1 + \phi_2)\sigma_z}$. Here $\sigma_x$ and $\sigma_z$ are the Pauli operators. The GHZ-symmetric states of three-qubits can be written as:\\
$\rho(p, q) = (\frac{2q}{\sqrt{3}}+ p)|GHZ_{+}\rangle\langle GHZ_{+}| \,+$
\begin{equation}\label{g1}
(\frac{2q}{\sqrt{3}}- p)|GHZ_{-}\rangle\langle GHZ_{-}| + (1- \frac{4q}{\sqrt{3}})\frac{\mathbf{1}}{8}
\end{equation}
where $|GHZ_{\pm}\rangle = \frac{|000\rangle \pm |111\rangle}{\sqrt{2}}$. The requirement $\rho(p, q)\geq 0$ gives the constraints: $-\frac{1}{4\sqrt{3}}\leq q \leq \frac{\sqrt{3}}{4}$ and
\begin{equation}\label{g8}
|p| \leq \frac{1}{8}+\frac{\sqrt{3}}{2} q.
\end{equation}
This family of states forms a triangle in the state space and includes not only GHZ states, but also the maximally mixed state $\frac{\textbf{1}}{8}$(located at the origin, see Fig.1). Any point inside that triangle represents a GHZ-symmetric state. The generalized werner states are found on the straight line $q = \frac{\sqrt{3}p}{2}$ connecting the origin with the $|GHZ_{+}\rangle$ state. A GHZ-symmetric state is fully separable iff it is in the polygon defined by the four corner points $(0,-\frac{1}{4\sqrt{3}})$, $(\frac{1}{8}, 0)$, $(0, \frac{\sqrt{3}}{4})$ and $(- \frac{1}{8}, 0)$ (yellow area in Fig.1). It is at most biseparable if and only if $|p| \leq \frac{3}{8}-\frac{\sqrt{3}}{2} q $( magenta area in Fig.1) . It is of $W$ type( grey area in Fig.1) if and only if $9216 p^4 + p^2 (-6768 + 17856 \sqrt{3} q - 34560 q^2 - 1024 \sqrt{3} q^3) \leq 1521 - 5148 \sqrt{3} q + 13536 q^2 + 2432 \sqrt{3} q^3 - 13056 q^4 - 3072 \sqrt{3} q^5$ and $|p| > \frac{3}{8}-\frac{\sqrt{3}}{2} q $.
\begin{figure}[htb]
\includegraphics[width=2.6in]{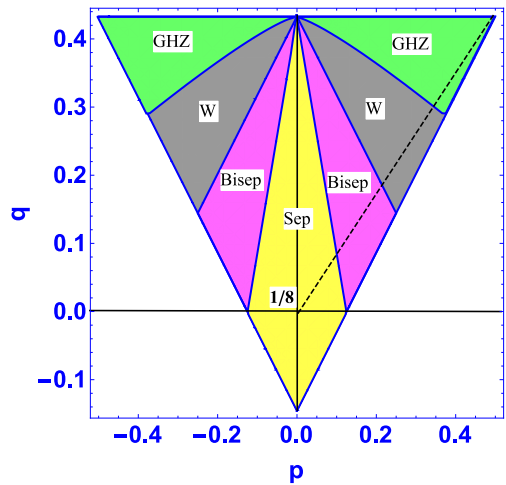}
\caption{\emph{The triangle of the GHZ-symmetric states for three qubits\cite{Elt}. The upper corners of the triangle are the standard GHZ states $|GHZ_{+}\rangle$ and $|GHZ_{-}\rangle$. Mixed state $\frac{\textbf{1}}{8}$ is located at the origin. The black dashed line represents the generalized Werner state. We have indicated different types of three qubit entanglement: GHZ(green), W(grey), Biseparable(magenta), Separable(yellow).  }}
\end{figure}
\subsection{B. Genuine multipartite concurrence ($C_{GM})$}
In order to facilitate the discussion of our results, we briefly describe the genuine multipartite concurrence, a measure of genuine multipartite entanglement defined as\cite{Zma}:$C_{GM}(|\psi\rangle):= \textmd{min}_j\sqrt{2 (1-\Pi_j(|\psi\rangle))}$ where $\Pi_j(|\psi\rangle)$ is the purity of the $jth$ bipartition of $|\psi\rangle$. The genuine multipartite concurrence of three qubit $X$ states has been evaluated in \cite{Has}. It is given by:
\begin{equation}\label{4v}
  C_{GM}=2\,\textmd{max}_i\{0,|z_i|-w_i\}
\end{equation}
with $w_i=\sum_{j\neq i}\sqrt{a_jb_j}$ where $a_j$, $b_j$ and $z_j(j=1,2,3,4)$ are the elements of the density matrix of tripartite X states:

 \[\begin{bmatrix}
       a_1 & 0    & 0    & 0   &  0   &  0    & 0    & z_1 \\
       0   & a_2  & 0    & 0   &  0   &  0    & z_2  &  0  \\
       0   & 0    & a_3  & 0   &  0   &  z_3  & 0    &  0  \\
       0   & 0    & 0    & a_4 &  z_4 &  0    & 0    &  0  \\
       0   & 0    & 0   &{z_4}^\ast &b_4 & 0  & 0   &    0   \\
       0   & 0    &{z_3}^\ast   &0 &  & b_3  & 0   &    0  \\
       0   &{z_2}^\ast    & 0  & 0 & 0 & 0 & b_2 & 0 \\
       {z_1}^\ast & 0 &0 & 0 & 0 & 0 & 0 & b_1 \\
 \end{bmatrix}\]
\subsection{C. Tripartite nonlocality}
In this section we provide a brief overview of the various notions of tripartite nonlocality and corresponding detectors of tripartite nonlocality for subsequent discussions. Consider a Bell-type experiment consisting of three space-like separated parties, Alice, Bob and Charlie. The measurement settings are denoted by $x$, $y$, $z$ $\in \{0,1\}$  and their outputs by $a$, $b$, $c$ $\in \{-1, 1\}$ for Alice, Bob and Charlie respectively. The experiment is thus characterized by the joint probability distribution(correlations) $p(abc|xyz)$. Now the correlations can exhibit different types of nonlocality. Any tripartite correlation $p(abc|xyz)$ is said to be local if it admits the following decomposition:
 \begin{equation}\label{g2}
    p(abc|xyz) = \sum_{\lambda}q_{\lambda}P_{\lambda}(a|x)P_{\lambda}(b|y)P_{\lambda}(c|z)
\end{equation}
for all $x$, $y$, $z$, $a$, $b$, $c$, where $0\leq q_{\lambda}\leq 1$ and $\sum_{\lambda}q_{\lambda}=1.$ $P_{\lambda}(a|x)$ is the conditional probability of getting outcome $a$ when the measurement setting is $x$ and $\lambda$ is the hidden variable; $P_{\lambda}(b|y)$ and $P_{\lambda}(c|z)$ are similarly defined. Otherwise they are standard nonlocal. We denote $L_3$ as the set of local correlations that can be produced classically using shared randomness. The local set $L_3$ was fully characterized by Pitowsky and Svozil\cite{Pit} and \'{S}liwa \cite{Sli}. It has $53856$ facets defining
$46$ different classes of inequalities that are inequivalent under relabeling of parties, inputs, and outputs\cite{Sli}. Violation of any of these facet inequalities guarantees standard nonlocality. A tripartie correlation is local if it satisfies all the $46$ facet inequalities. Inequality 2(we follow \'{S}liwa's numbering) is the Mermin inequality\cite{Mer}:
 \begin{equation}\label{g3}
 M = |\langle A_1B_0C_0\rangle + \langle A_0B_1C_0\rangle + \langle A_0B_0C_1\rangle - \langle A_1B_1C_1 \rangle| \leq 2
\end{equation}
Note that it is possible to violate Mermin inequality maximally(i.e., $M = 4$) using $|GHZ_{\pm}\rangle$.\\
However, in tripartite scenario, Svetlichny\cite{Svt} showed that there exist certain quantum correlations which can exhibit an even stronger form of nonlocality. Such type of correlations cannot be decomposed in the following form:
$$P(abc|xyz) =\sum_{\lambda}q_{\lambda}P_{\lambda}(ab|xy ) P_{\lambda}(c|z)+$$
\begin{equation}\label{g4}
    \sum_{\mu} q_{\mu}P_{\mu}(ac|xz) P_{\mu}(b|y ) +\sum_{\nu} q_{\nu}P_{\nu}(bc|yz) P_{\nu}(a|x);
\end{equation}
Here $0\,\leq\,q_{\lambda},\,q_{\mu},q_{\nu}\,\leq\,1$ and $\sum_{\lambda}q_{\lambda}+\sum_{\mu}q_{\mu}+\sum_{\nu}q_{\nu}=1.$
The above form of correlations are not fully local as in Eq. (\ref{g2}), nonlocal correlations are present only between two particles (the two particles that are nonlocally correlated can change in different runs of the experiment) while they are only locally correlated with the third. If a correlation $P(abc|xyz)$ cannot be written in this form then such a correlation is said to exhibit genuine tripartite nonlocality. In \cite{Ban}, this type of nonlocality is referred to as Svetlichny nonlocality. Focusing on these form of correlations(Eq.(\ref{g4})), Svetlichny designed a tripartite Bell type inequality (known as Svetlichny inequality):
\begin{equation}\label{g5}
 S\leq4.
\end{equation}
where $ S\,=\, |\langle A_0B_0C_0\rangle+\langle A_1B_0C_0\rangle-\langle A_0B_1C_0\rangle +$ $$\langle A_1B_1C_0 \rangle   +\langle A_0B_0C_1\rangle-\langle A_1B_0C_1\rangle+\langle A_0B_1C_1\rangle +\langle A_1B_1C_1\rangle|.$$
Thus violation of such inequality implies the presence of genuine tripartite nonlocality, implying in turn the presence of genuine tripartite entanglement. This inequality(\ref{g5}) is violated by GHZ and W states\cite{Gho,Ajo,Lav}.\\
While Svetlichny's notion of genuine multipartite nonlocality is often referred to in the literature, it has certain drawbacks. As has been pointed out in \cite{Ban,Gal}, Svetlichny's notion of genuine tripartite nonlocality is so general that no restrictions were imposed on the bipartite correlations used in Eq. (\ref{g4}). They are allowed to display arbitrary correlations in the sense that there may be one-way or both way signaling between a pair of parties or both the parties may perform simultaneous measurements. As a result, grandfather-type paradoxes arise \cite{Ban} and inconsistency from an operational viewpoint appears\cite{Gal}. Moreover it is found that there exist some genuine nonlocal correlations which satisfy this inequality\cite{Ban,Gal,Kau}. In order to remove this sort of ambiguity, Bancal et al.\cite{Ban}, introduced a simpler definition of genuine tripartite nonlocality which is based on no-signaling principle, in which the correlations are no-signaling for all observers. Suppose $P(abc|xyz)$ be the tripartite correlation satisfying Eq.(\ref{g4}) with no-signaling criteria imposed on the bipartite correlations terms, i.e.,
\begin{equation}\label{4ii}
   P_{\lambda}(a|x)=\sum_b P_{\lambda}(ab|xy )\, ~~\forall\, a,\, x,\, y,
\end{equation}
\begin{equation}\label{4iii}
  P_{\lambda}(b|y)=\sum_a P_{\lambda}(ab|xy ) \, ~~\forall \,b, x,\, y.
\end{equation}
and similarly for the other bipartite correlation terms $P_{\mu}(ac|xz)$ and $P_{\nu}(bc|yz).$ The above form of correlations are called $NS_2$ local. Otherwise, we say that they are genuinely $3-$way NS nonlocal($NS_2$ nonlocal). In \cite{Ban}, 185 Bell-type inequalities are given which constitute the full class of facets of $NS_2$ local polytope. Violation of any of these facets(Bell-type inequalities) guarantees $NS_2$ nonlocality. Svetlichny inequality constitute the 185-th class. Throughout the paper, we use this notion of nonlocality as genuine tripartite nonlocality.
\section{III.STANDARD NONLOCALITY OF GHZ-SYMMETRIC STATES}
We have already discussed in introduction that all tripartite pure  entangle state exhibit standard nonlocality but this relation does not hold for mixed states. From that point of view and also from experimental perspectives, characterization of mixed states on the basis of their ability to generate nonlocal correlations is far more interesting compared to that of pure states. As already discussed before, we aim to characterize the state parameters of the mixed class of GHZ-symmetric three-qubit states on the basis of their nonlocal nature. In this section we not only classify this class on the basis of standard nonlocality but also derived the necessary and sufficient condition of detecting standard nonlocality.
\subsection{A. Maximum violation of Mermin inequality}
We have already mentioned that standard nonlocality of correlations can be detected if the correlations violate atleast one of the $46$ inequivalent facet inequalities characterizing the local set($L_3$). Among the $46$ inequivalent facet inequalities, Mermin inequality is most frequently used. In \cite{Chi}, they gave a sufficient criterion to violate  Mermin inequality for pure three qubit states. Here we find a necessary and sufficient condition to  obtain a violation of Mermin inequality for three qubit GHZ-symmetric states. For this class of tripartite mixed entangled states the maximum  value of $M$(Eq.(\ref{g3})) with respect to projective measurement is given by $8|p| $(see Appendix). Then the Mermin inequality in  Eq.(\ref{g3}) becomes:
\begin{equation}\label{g7}
  M_{max} =  8|p| \leq 2.
\end{equation}
Hence $\rho(p, q)$ violates Mermin inequality if and only if $|p| > \frac{1}{4}.$ Due to this restriction on $p$, together with state constraints Eq.(\ref{g8}), the other state parameter $q$ gets restricted: $q > \frac{1}{4\sqrt{3}}$. So standard nonlocality of any three qubit GHZ-symmetric states with $|p| > \frac{1}{4}$ and $q > \frac{1}{4\sqrt{3}}$ is guaranteed via violation of Mermin inequality(see Fig.2(a)).
\subsection{B.Efficiency of $15$-th facet inequality over Mermin inequality}
Mermin inequality, discussed in the last section, is not the most efficient detector of standard nonlocality. In this section it is argued that there exists another facet inequality which can be considered as a better tool for detecting standard nonlocality compared to use of Mermin inequality for doing the same. It is observed that the $15$-th facet can be considered as an inequality which is more efficient than Mermin inequality.

The $15$-th facet inequality is given by\cite{Sli}:
\begin{equation}\label{g9}
  L \leq 4.
\end{equation}
where $L = |2\langle A_0B_0\rangle+ 2\langle A_1B_0\rangle+\langle A_0C_0\rangle +\langle A_1C_0 \rangle -2\langle B_0C_0\rangle+\langle A_0B_1C_0\rangle-\langle A_1B_1C_0\rangle +\langle A_0C_1\rangle+ \langle A_1C_1\rangle - 2\langle B_0C_1\rangle-\langle A_0B_1C_1\rangle+\langle A_1B_1C_1\rangle|.$
The maximum value of $L$ for three qubit GHZ-symmetric states with respect to projective measurement is given by $\max[\frac{8(9|p^3|-8\sqrt{3}|q^3|)}{9p^2-12q^2}, -16\sqrt{3}q|]$ (see Appendix). Using this, $15$-th facet inequality(Eq.(\ref{g9})) gets modified as
\begin{equation}\label{g10}
 L_{max} = \max[\frac{8(9|p^3|-8\sqrt{3}|q^3|)}{9p^2-12q^2}, -16\sqrt{3}q|] \leq 4
\end{equation}
As, $-16\sqrt{3}q\leq4$ for $-\frac{1}{4\sqrt{3}}\leq q \leq \frac{\sqrt{3}}{4}$, so $15$-th facet inequality is violated only when $\frac{8(9|p^3|-8\sqrt{3}|q^3|)}{9p^2-12q^2} > 4.$ Using this relation and Eq.(\ref{g8}), we have $q>\frac{3}{148}(8+3\sqrt{3})$. It follows that for every $q>\frac{3}{148}(8+3\sqrt{3})$ there is atleast one $p$ for which the GHZ-symmetric states violate $15$-th facet inequality. In Fig.2(b) we have plotted the range of the state parameters for which nonlocality is observed  via the violation of $15$-th facet inequality. We have already discussed that a GHZ-symmetric state do not violate Mermin inequality if and only if $|p|\leq \frac{1}{4}$. Now this restriction, when imposed on $L_{max}>4$ gives atleast one $q>\frac{1}{16}(\sqrt{15}+\sqrt{3})$ for all nonzero $p$. Hence there exists a region for $|p|\leq \frac{1}{4}$  where $15$-th facet inequality helps us to reveal standard nonlocality unlike Mermin inequality where the same is revealed only for $|p|>\frac{1}{4}$. For example let us consider the GHZ-symmetric states with $p=0.2$ and $q\in [-\frac{1}{4\sqrt{3}}, \frac{\sqrt{3}}{4}]$. This state does not violate Mermin inequality for any value of $q$ but the same state violates $15$-th facet inequality for $q>0.37861$.  This in turn points out that $15$-th facet inequality is more efficient than the Mermin inequality over some restricted range of state parameters.
\subsection{C. Necessary and sufficient detection criteria of Standard nonlocality}
By comparing the criteria necessary and sufficient for violation of each of the remaining $44$ inequivalent facet inequalities(following procedure similar to that used for Mermin inequality, see Appendix) with that of Mermin inequality and the $15$-th facet inequality, we have observed that region of standard nonlocality, as detected by any of the remaining $44$ inequivalent facet inequalities forms subset of the region of standard nonlocality of Mermin and the $15$-th facet inequality. So these two inequalities are the most efficient to detect standard nonlocality of this class of states. This in turn points out that the optimal region of standard nonlocality of GHZ-symmetric states is provided by the union of regions of standard nonlocality detected by Mermin and the $15$-th facet inequality(see Fig.2(c)). So in totality the restricted state conditions for revealing standard nonlocality are given by:\\
$(i)$\,\, $4|p| > 1,\,\,q > \frac{1}{4\sqrt{3}}$
and
\begin{equation}\label{g69}
   (ii)\,\, \frac{8(9|p^3|-8\sqrt{3}|q^3|)}{9p^2-12q^2}> 4\,,\,q>\frac{3}{148}(8+3\sqrt{3}).
\end{equation}
A state exhibits standard nonlocality under projective measurements if and only if it satisfies atleast one of the two sets of conditions($(i)$ or $(ii)$ of Eq.(\ref{g69})).
\begin{figure}[htb]
\includegraphics[width=1.6in]{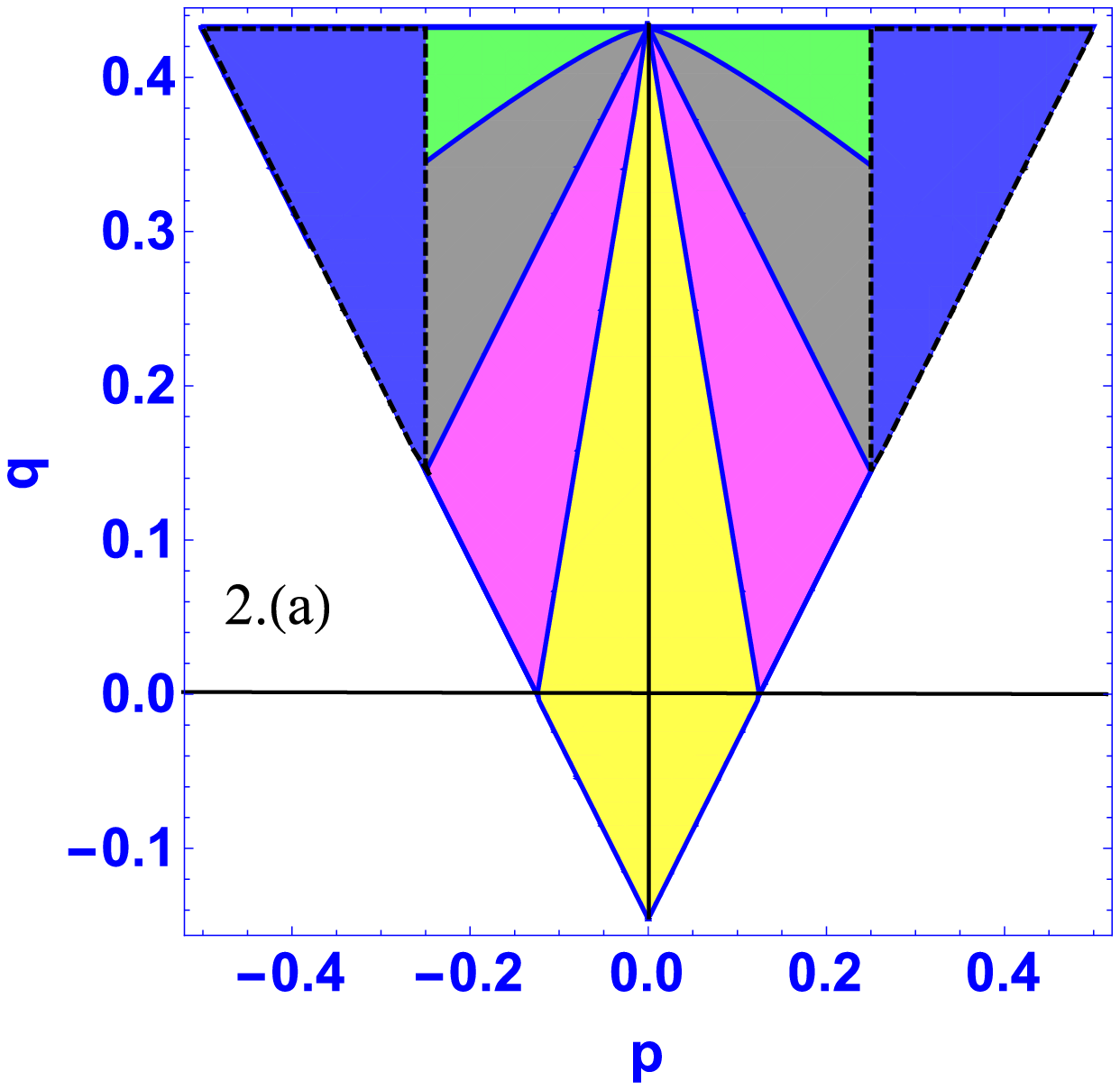}
\includegraphics[width=1.6in]{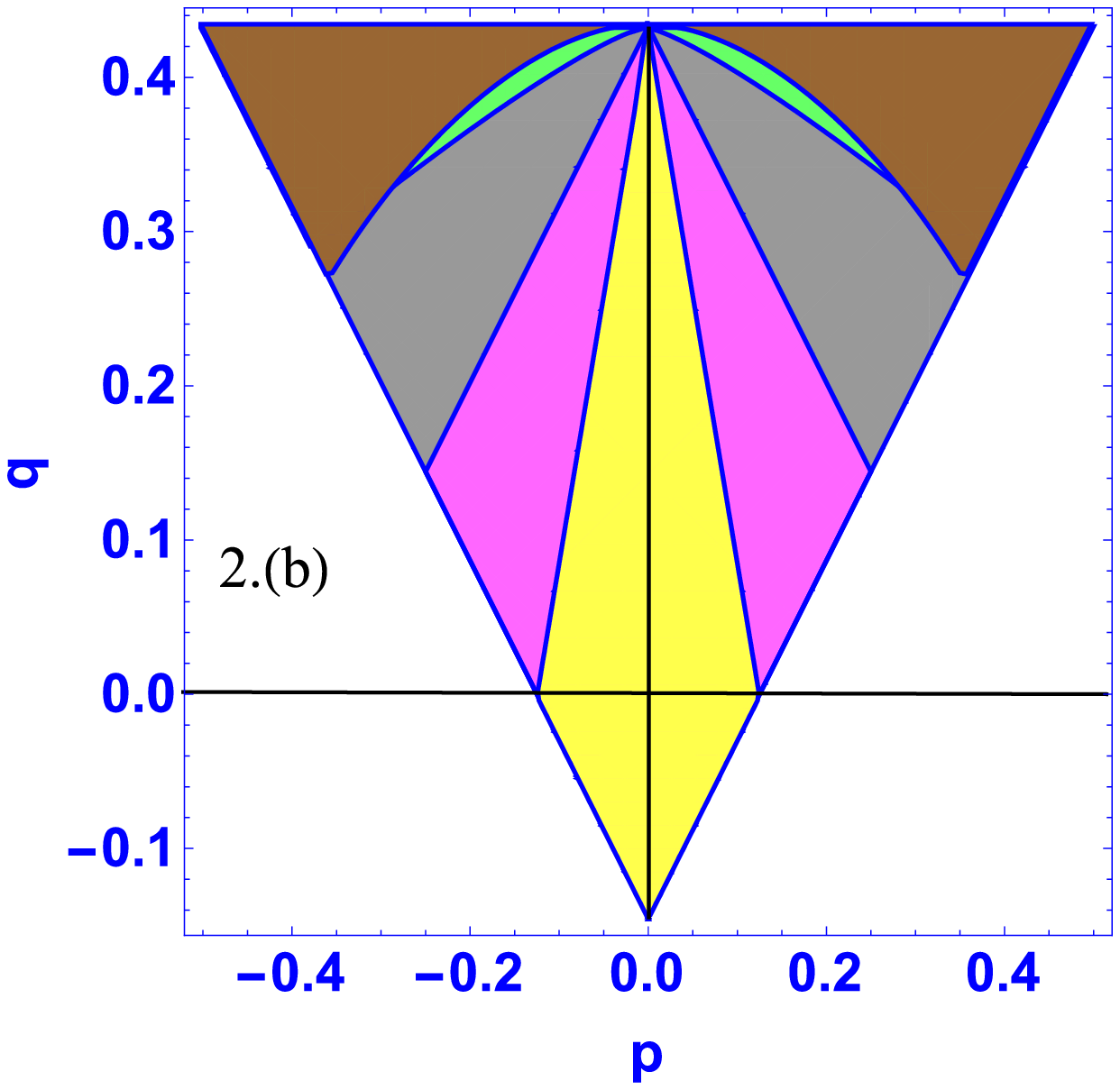}\\
\includegraphics[width=2.6in]{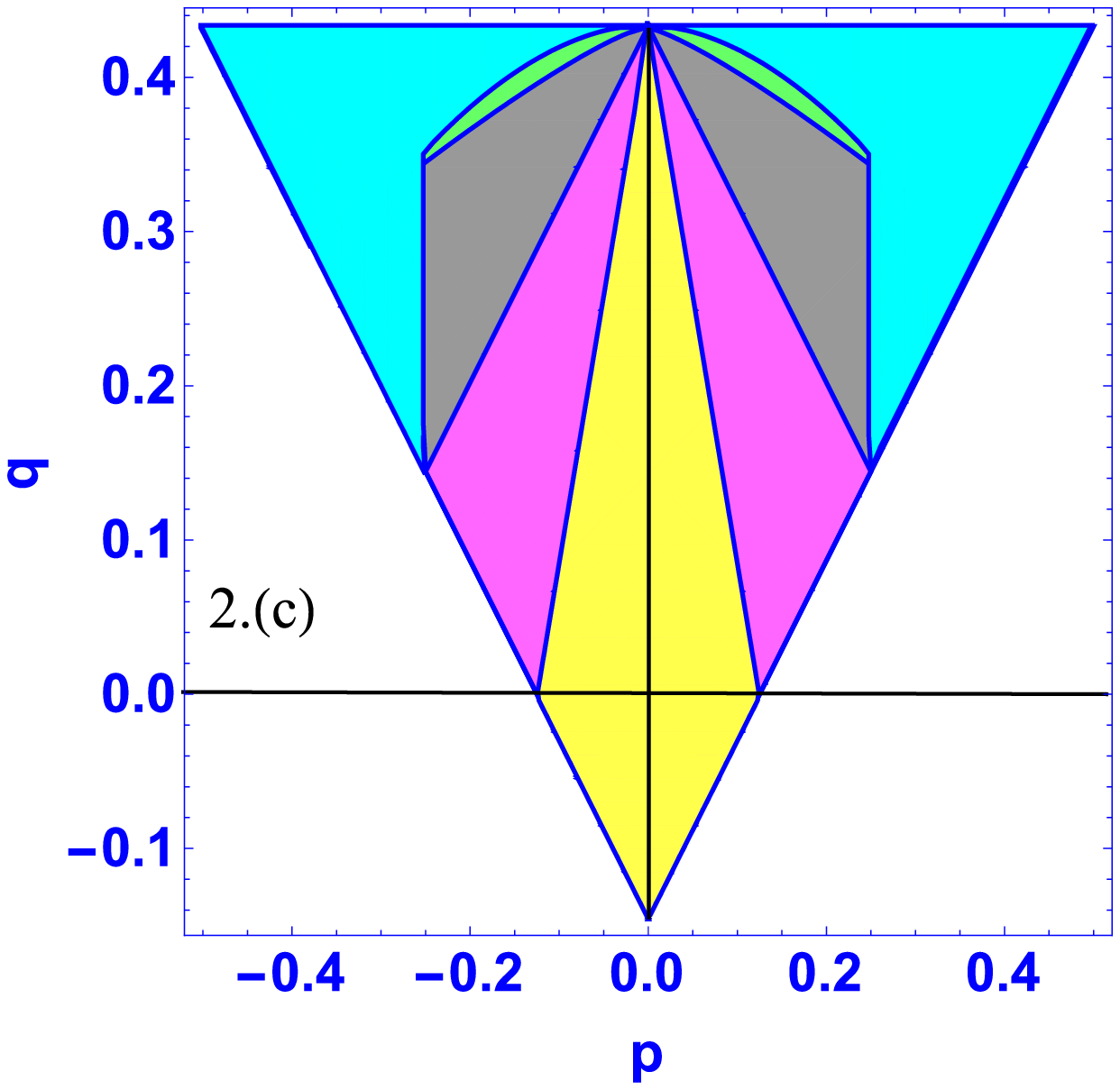}
\caption{\emph{(a) The blue areas represent the nonlocal region obtained via the violation of Mermin inequality(Eq.\ref{g3}). (b)  Regions of violation of $15$-th inequality are given by the brown regions. Clearly, the regions restricted by $p\leq\frac{1}{4}$ in the brown regions indicate the areas where $15$-th facet inequality(Eq.(\ref{g9})) emerges as a more efficient tool over Mermin inequality(Eq.\ref{g3}) for revealing nonlocal nature of GHZ-symmetric states. (c) The cyan areas give the optimal region of standard nonlocality of GHZ-symmetric states. The states characterized by the state parameters lying in this region, when shared between Alice, Bob and Charlie do not admit any local hidden variable model.}}
\end{figure}
\section{IV. GENUINE NONLOCALITY OF GHZ-SYMMETRIC STATES}
Genuine nonlocality is the strongest form of nonlocality. So for a tripartite correlation it is natural to ask whether all three parties are nonlocally correlated. Such correlations play an important role in quantum information theory, phase transitions and in the study of many-body systems\cite{Zol}. Also, the presence of genuine nonlocality implies the presence of genuine entanglement. So after discussing about standard nonlocality, it becomes interesting to explore about genuine nonlocality of this class of states. To be specific, in this section we have derived necessary and sufficient criteria for detecting genuine nonlocality. \\
\subsection{A. Maximum violation of Svetlichny inequality}
As we have discussed before, if we consider that all correlations between the observers are no-signalling, then  the set of $185$ facet inequalities act as a necessary and sufficient condition for detecting genuine tripartite nonlocality. Among all of them, Svetlichny inequality is frequently used for the detection of genuine tripartite nonlocality. In \cite{Gho}, necessary and sufficient criteria for maximal violation of Svetlichny inequality are derived for some classes of tripartite pure states. Here we have derived the same for the class of GHZ-symmetric states. For this class of tripartite mixed entangled states, the maximum value of S(Eq.(\ref{g5})) with respect to projective measurement is given by $8\sqrt{2}|p|$(see Appendix).
Thus Eq.(\ref{g5}) gives,
\begin{equation}\label{g11}
   S_{max} = 8\sqrt{2}|p| \leq 4.
\end{equation}
Hence $\rho(p, q)$ violates Svetlichny inequality if and only if $|p|>\frac{1}{2\sqrt{2}}$. Using this relation and Eq.(\ref{g8}), we have $q>\frac{1}{\sqrt{3}}(\frac{1}{\sqrt{2}}-\frac{1}{4})$. So Svetlichny nonlocality is revealed for three qubit GHZ-symmetric states if and only if  the relation $|p|>\frac{1}{2\sqrt{2}}$ and $q>\frac{1}{\sqrt{3}}(\frac{1}{\sqrt{2}}-\frac{1}{4})$ holds. In Fig.3(a) we present the range of the state parameters of the three-qubit GHZ-symmetric states for which Svetlichny nonlocality is observed. \\
\subsection{B. Efficiency of $99$-th facet inequality over Svetlichny inequality}
As we have mentioned in Section II, the newly introduced weaker definition of genuine nonlocality(genuine 3-way NS nonlocality) gives advantage over Svetlichny's definition of genuine nonlocalty.  So after completing the analysis of genuine nonlocality with respect to Svetlichny inequality, we search for an inequality which can be considered more efficient than Svetlichny inequality. In \cite{Kau}, we have shown that for detecting genuine nonlocality of some classes of tripartite pure entangled states, $99$-th facet inequality is more efficient compared to Svetlichny inequality.  Here also, for the class of GHZ-symmetric states, $99$-th facet inequality emerges to be more powerful tool for detecting genuine nonlocality for some subclasses.  The $99$-th facet inequality is given by:
\begin{equation}\label{g12}
  NS \leq 3.
\end{equation}
where $NS$=$|\langle A_1B_1\rangle +   \langle A_0B_0C_0\rangle +   \langle B_1C_0\rangle +   \langle A_1C_1\rangle -  \langle A_0B_0C_1\rangle|.$ If projective measurement is considered, the maximum value of $NS$ is given by $\frac{4q}{\sqrt{3}} + 2\sqrt{\frac{16 q^2}{3}+ 4p^2 }$(see Appendix). Thus $99$-th facet inequality in Eq.(\ref{g12}) becomes,
\begin{equation}\label{g13}
 NS_{max} = \frac{4q}{\sqrt{3}} + 2\sqrt{\frac{16 q^2}{3}+ 4p^2 } \leq 3.
\end{equation}
Hence $99$-th facet inequality is violated if and only if $\frac{4q}{\sqrt{3}} + 2\sqrt{\frac{16 q^2}{3}+ 4p^2 } > 3.$ Using this along with the state constraints(Eq.\ref{g8}), we have $q>\frac{1}{28}(8\sqrt{5}-5\sqrt{3}).$ Thus $NS_2$ nonlocality is observed if $\frac{4q}{\sqrt{3}} + 2\sqrt{\frac{16 q^2}{3}+ 4p^2 } > 3$ and $q>\frac{1}{28}(8\sqrt{5}-5\sqrt{3}).$ Hence for every $q>\frac{1}{28}(8\sqrt{5}-5\sqrt{3})$ there exists atleast one GHZ-symmetric state which is $NS_2$ nonlocal. We have already observed that any state restricted by $|p| \leq \frac{1}{2\sqrt{2}}$ fails to violate Svetlichny inequality. Now this restriction, when imposed on $NS_{max} > 3$ gives atleast one $q>\frac{1}{4}(\sqrt{10}-\sqrt{3})$ for all nonzero $p$. Hence we get a subclass of GHZ-symmetric states restricted by $q>\frac{1}{4}(\sqrt{10}-\sqrt{3})$ and $NS_{max}>3$ which is genuinely nonlocal even when $|p| \leq \frac{1}{2\sqrt{2}}$. This in turn points out efficiency of $99$-th facet inequality over Svetlichny inequality.
\subsection{C. Necessary and sufficient criteria for detecting Genuine nonlocality} A detailed comparison of the criteria required for violation of each of the remaining $183$ facets(following same procedure as that for Mermin inequality) with that of Svetlichny and $99$-th facet points out the fact that these two inequalities($99$-th facet inequality and Svetlichny inequality) are the most efficient detectors of genuine nonlocality. This in turn points out the fact that the optimal region of genuine nonlocality is given by:
$(i)$\,\, $|p|>\frac{1}{2\sqrt{2}}$,\,\,$q>\frac{1}{\sqrt{3}}(\frac{1}{\sqrt{2}}-\frac{1}{4})$
and
\begin{equation}\label{g19}
   (ii)\,\, \frac{4q}{\sqrt{3}} + 2\sqrt{\frac{16 q^2}{3}+ 4p^2 } > 3\,,\,q>\frac{1}{28}(8\sqrt{5}-5\sqrt{3}).
\end{equation}
Genuine nonlocality of any state, upto projective measurements is guaranteed  if and only if it satisfies atleast one of the two possible sets of conditions($(i)$ or $(ii)$ of Eq.(\ref{g19})).
\begin{figure}[htb]
\includegraphics[width=1.6in]{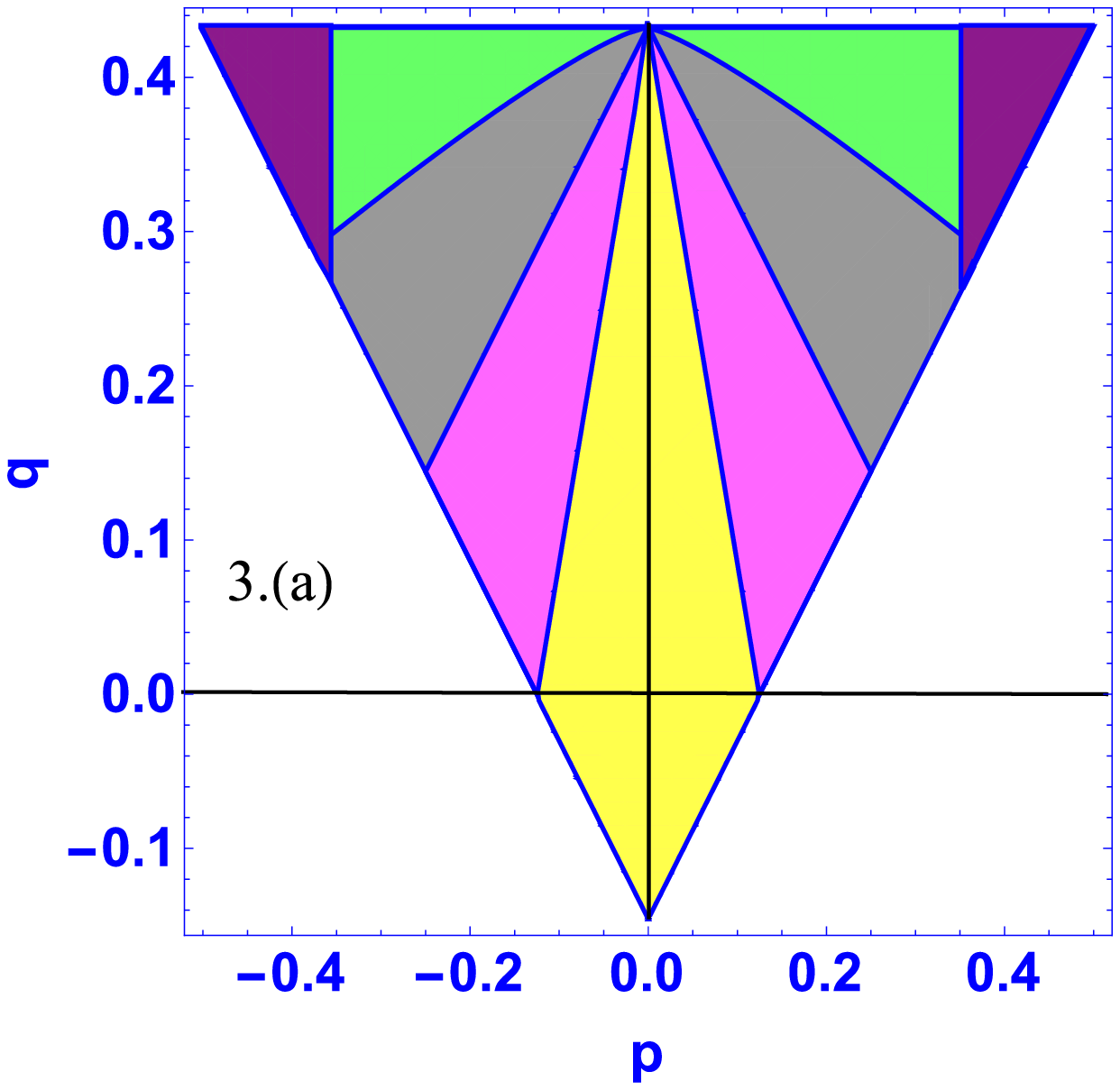}
\includegraphics[width=1.6in]{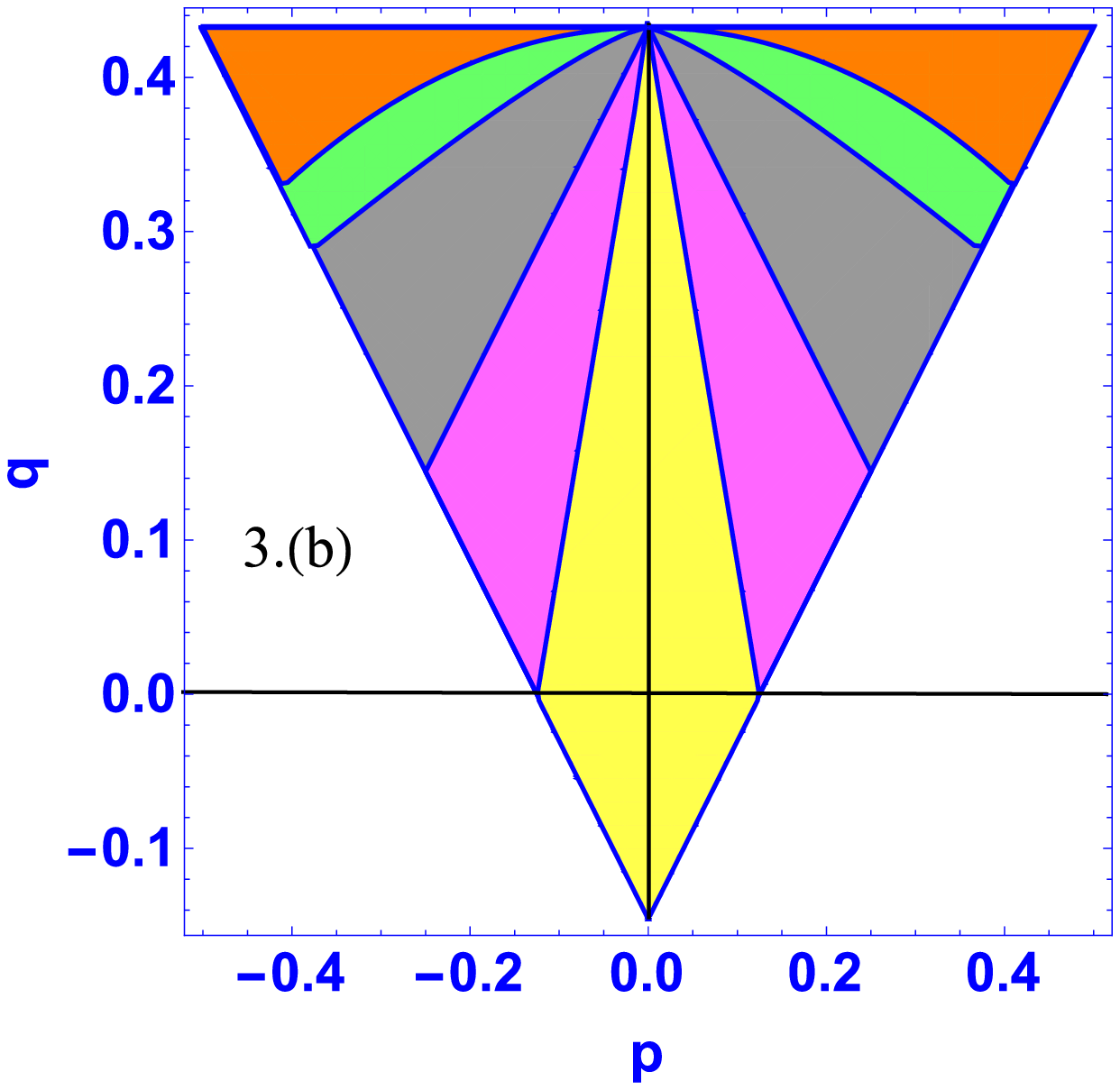}\\
\includegraphics[width=2.5in]{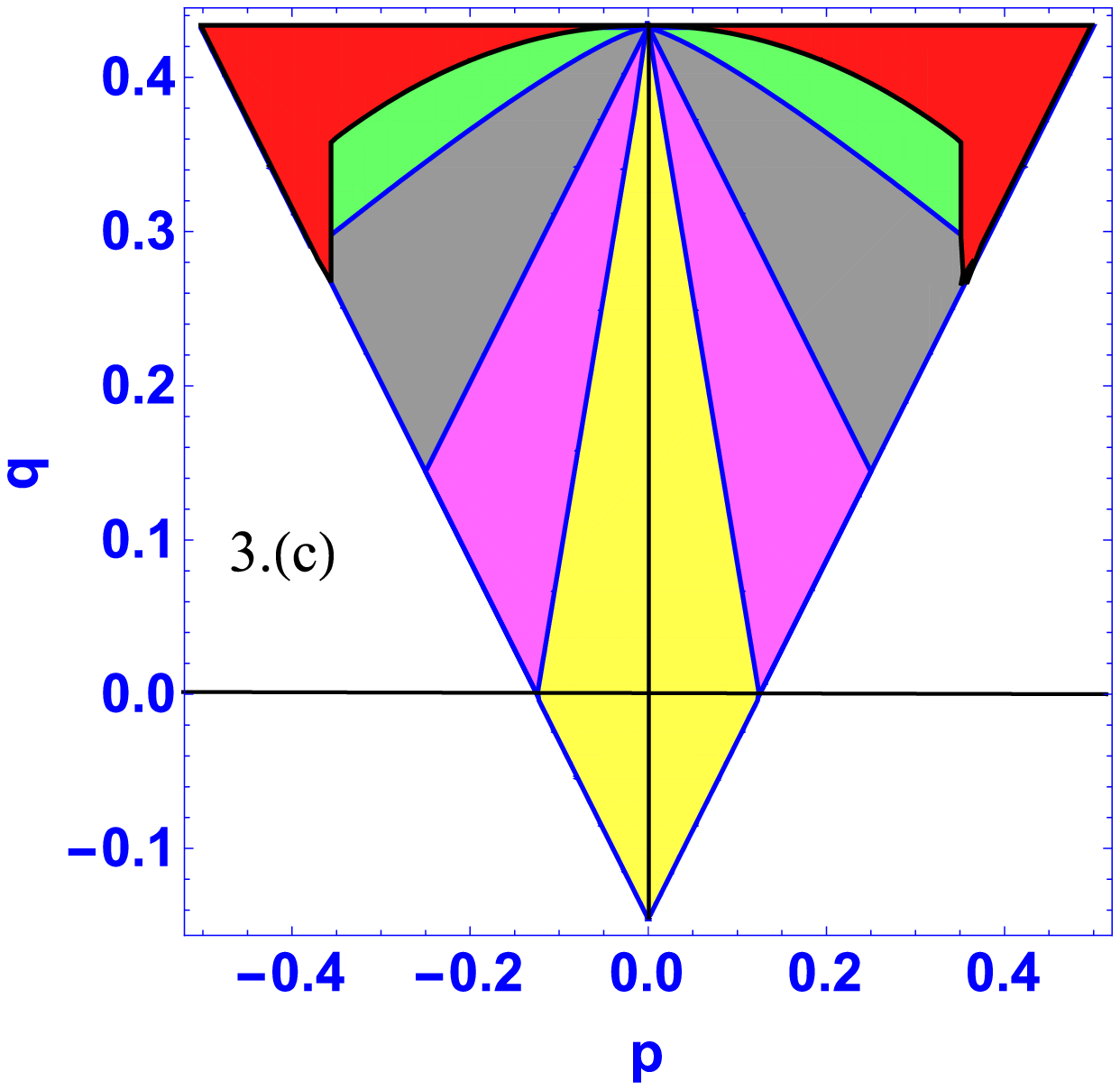}
\caption{\emph{(a)The purple areas give the restricted region of state parameters for which genuine nonlocality of the corresponding states is guaranteed by violation of Svetlichny inequality(Eq.\ref{g5}). (b) Analogously, the orange regions represent the areas where genuine nonlocality is observed due to violation of $99$-th facet inequality(Eq.(\ref{g12})).  Now the regions restricted by $|p| \leq \frac{1}{2\sqrt{2}}$ in the orange regions give the the areas where the  $99$-th facet serves as a better tool to exploit genuine nonlocality of  GHZ-symmetric states compared to Svetlichny inequality.(c) The optimal region of nonlocality of GHZ-symmetric states is given by the red regions. }}
\end{figure}
\section{V. RELATION BETWEEN ENTANGLEMENT AND NONLOCALITY}
Entanglement of any state is necessary for nonlocality of the state. So after completing classification of GHZ-symmetric states with respect to different form of nonlocality, we proceed to establish the relationship between nonlocality and entanglement of this class.
\subsection{A. Relation between Biseparable entanglement and Standard nonlocality }
Biseparable entanglement of a tripartite quantum state is necessary to produce standard nonlocality. Their relationship has been analyzed in \cite{Syu} for three qubit pure states where it is shown that  biseparable entanglement of tripartite pure quantum states also turns out to be sufficient to exhibit standard nonlocality.  Here we analyze whether it is sufficient for this class of tripartite mixed quantum states to obtain standard nonlocality.
The criterion of biseparability of this class of states is\cite{Elt,Buc}: $|p| \leq \frac{3}{8}-\frac{\sqrt{3}}{2} q $. Interestingly, no biseparable GHZ-Symmetric state can reveal standard nonlocality. We present our argument below.\\
 We have already discussed that to detect standard nonlocality, Mermin and the $15$-th facet inequality are the most efficient inequalities. In order to violate Mermin inequality the state parameters should satisfy $|p| > \frac{1}{4}$ and $q > \frac{1}{4\sqrt{3}}$. However $|p| > \frac{1}{4}$, along with the biseparability criterion gives $q\leq\frac{1}{4\sqrt{3}}$. This contradicts the required criterion for violation of Mermin: $q>\frac{1}{4\sqrt{3}}.$ So violation of Mermin inequality is impossible. Now we consider the $15$-th facet inequality. Using the biseparability criterion, we get $L_{max}\leq \frac{8(9(\frac{3}{8}-\frac{\sqrt{3q}}{2})^3-8\sqrt{3}|q^3|)}{9((\frac{3}{8}-\frac{\sqrt{3q}}{2})^2-12q^2)}$(say, $f$) where $f\leq 4$ and that makes violation of $15$-th inequality impossible by any biseparable state belonging to this class. Hence no biseparable state is capable of showing standard nonlocality. \\
  \begin{figure}[htb]
\includegraphics[width=2.5in]{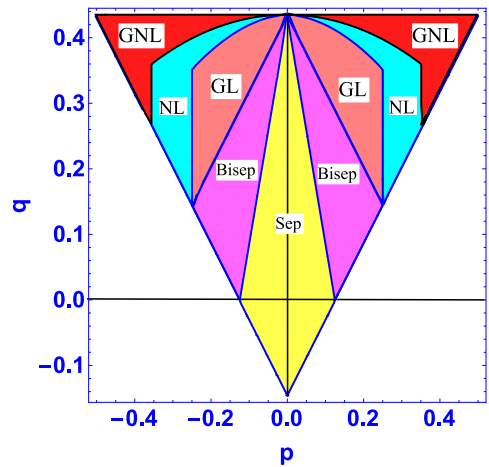}
\caption{\emph{The figure gives the nonlocality classification of three qubit GHZ-symmetric states $\rho(p, q)$. The red regions give the optimal area where genuine nonlocality(GNL) is revealed with respect to projective measurement for GHZ-symmetric states. The cyan regions indicate the optimal area where standard nonlocality(NL) is revealed(except genuine nonlocality). As genuine nonlocality also implies standard nonlocality so red regions also give the region of standard nonlocality. Genuinely entangled but local states(GL) are represented by pink regions. Clearly, no nolocal region lies within biseparable region(Magenta). }}
\end{figure}
\subsection{B. Relation between Genuine entanglement and Standard nonlocality }
In general for any tripartite state,  genuine entanglement is necessary to reveal genuine nonlocality.  So for GHZ-symmetric states, as argued in the last section, genuine entanglement is necessary to reveal even the weaker notion of standard nonlocality. However one cannot claim it to be a sufficient criterion for revealing standard nonlocality for this class of states. We proceed with our argument below. For that we first consider genuinely entangled states. Such states are restricted by\cite{Elt,Buc}:
\begin{equation}\label{g20}
    |p| > \frac{3}{8}-\frac{\sqrt{3}}{2}q
\end{equation}
As we have discussed earlier the locality criteria are: $$4|p| \leq 1$$ and
\begin{equation}\label{g39}
 \frac{8(9|p^3|-8\sqrt{3}|q^3|)}{9p^2-12q^2}\leq 4
\end{equation}
Clearly the conditions(Eq.(\ref{g20}) and Eq.(\ref{g39})) are feasible with the restricted range of the parameter $q$ given by :$\frac{1}{4\sqrt{3}}\leq q \leq \frac{\sqrt{3}}{4}$. This in turn proves the existence of genuinely entangled local states(see the pink region of Fig.4). So any GHZ-symmetric state is genuinely entangled but local if it satisfies Eq.(\ref{g20}) and Eq.(\ref{g39}).  So strongest form of entanglement i.e., genuine entanglement turns out to be insufficient to generate even the weaker form of nonlocality i.e., standard nonlocality. Hence we are able to present a class of genuinely entangled three qubit states which does not violate a complete set of facet inequalities for standard nonlocality. Recently a similar type of result has been presented in \cite{Bow}, for some other class of states.\\
In this context it will be interesting to study variation of standard nonlocality with the amount of genuine entanglement. Since Mermin and $15$-th facets are the most efficient bell inequalities to detect standard nonlocality, so now we deal with the variation of violation of these facet inequalities with the amount  genuine entanglement($C_{GM}^{\rho(p, q)}$). Since the three-qubit GHZ-symmetric states belong to the class of tripartite X states, their amount of entanglement can be measured by Eq.(\ref{4v}). So \begin{equation}\label{g14}
C_{GM}^{\rho(p, q)} = 2 |p| - \frac{3}{4} + \sqrt{3}q.
\end{equation}
For the state $\rho(p, q)$, one has $M_{max} = 4(C_{GM}^{\rho(p, q)} + \frac{3}{4}-\sqrt{3}q)$. Hence, GHZ-symmetric states violate Mermin inequality if $(C_{GM}^{\rho(p, q)} + \frac{3}{4}-\sqrt{3}q)>\frac{1}{2}$. As we have proven in Section.III, GHZ-symmetric states with $|p|>\frac{1}{4}$ and $q > \frac{1}{4\sqrt{3}}$ violate Mermin inequality. For this subclass of GHZ-symmetric states $C_{GM}^{\rho(p, q)} > 0$ as $|p|>\frac{1}{4}$ and $q > \frac{1}{4\sqrt{3}}$. This subclass always violates Mermin inequality and the amount of violation(i.e.,$M_{max}-2$) increases monotonically with $C_{GM}^{\rho(p, q)}$ for any fixed value of $q$. Also for each $C_{GM}^{\rho(p, q)} > 0$ there is a GHZ-symmetric state(i.e., a value of $q$)which violates Mermin inequality(See Fig.5.(a)). Similarly, using Eq.(\ref{g14}), we have $L_{max} = \frac{8(-8\sqrt{3}q^3+\frac{9}{8}(C_{GM}^{\rho(p, q)} + \frac{3}{4}-\sqrt{3}q)^3)}{-12q^2+ \frac{9}{4}(C_{GM}^{\rho(p, q)} + \frac{3}{4}-\sqrt{3}q)^2)}$, which increases monotonically with $C_{GM}^{\rho(p, q)}$ for any fixed value of $q$. GHZ-symmetric states violate $15$-th facet inequality if and only if $\frac{2(-8\sqrt{3}q^3+\frac{9}{8}(C_{GM}^{\rho(p, q)} + \frac{3}{4}-\sqrt{3}q)^3)}{-12q^2+ \frac{9}{4}(C_{GM}^{\rho(p, q)} + \frac{3}{4}-\sqrt{3}q)^2)}>1$. Clearly, for each value of $C_{GM}^{\rho(p, q)}$ there is a GHZ-symmetric state with $q>\frac{3}{148}(8+3\sqrt{3})$ which violates $15$-th facet inequality. These are also shown in Fig.5.(b)
 \begin{figure}[htb]
\includegraphics[width=1.65in]{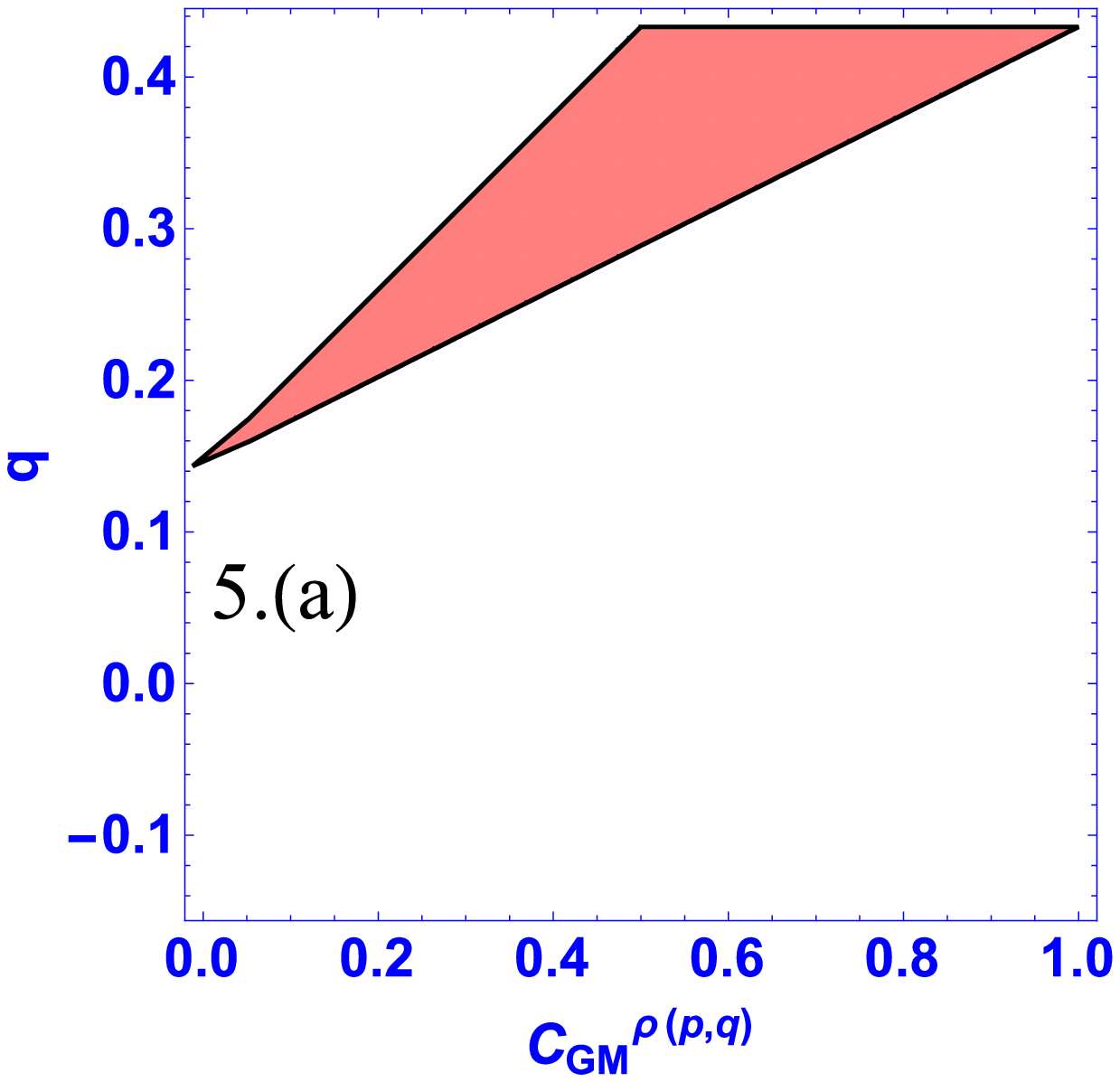}
\includegraphics[width=1.65in]{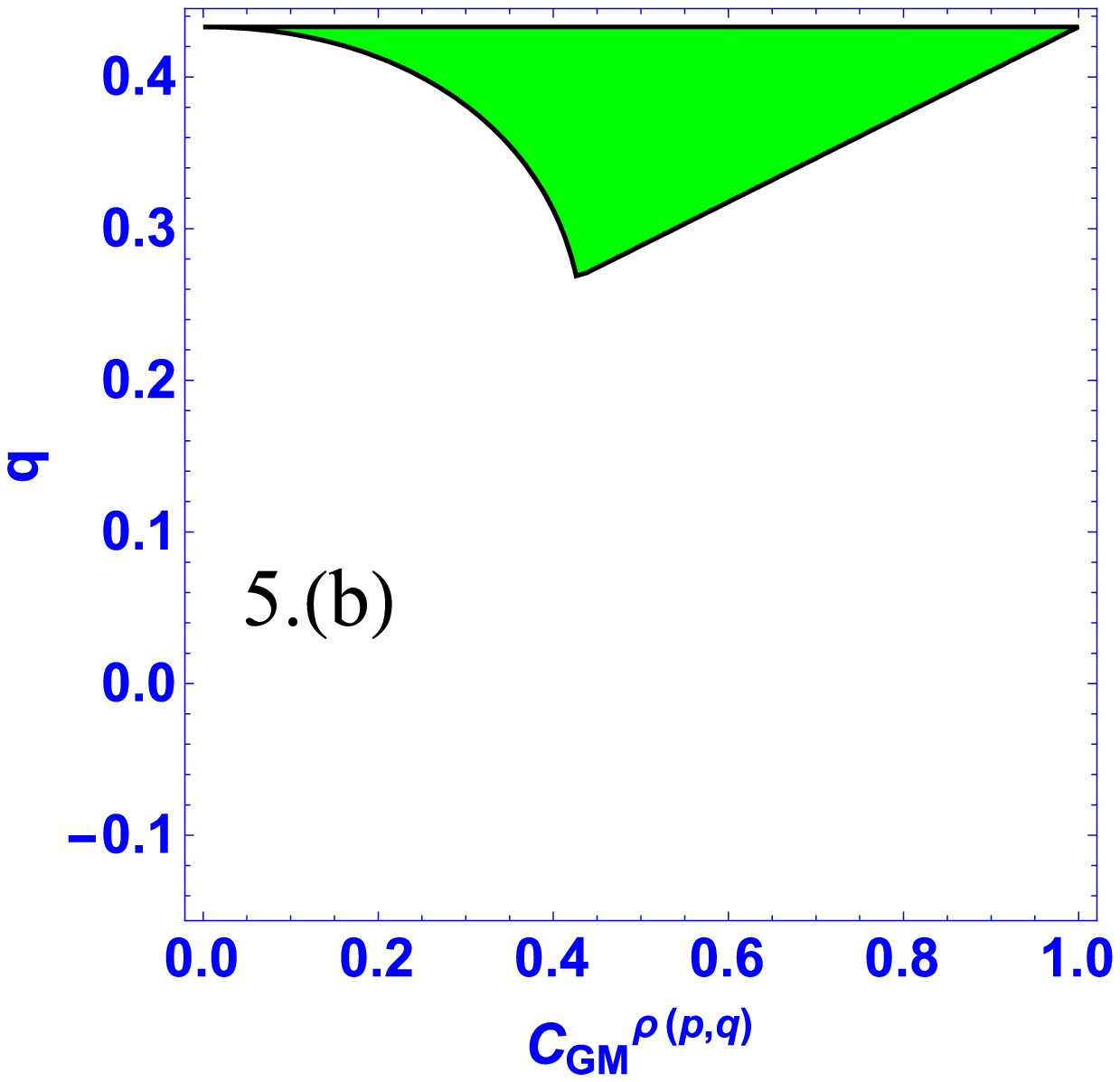}
\caption{\emph{Both of these figures depict variation of $C_{GM}^{\rho(p, q)}$ with state parameter $q$ for standard nonlocal states. Precisely, in Fig.(a) and Fig.(b), we have considered $C_{GM}^{\rho(p, q)}$ of the states whose standard nonlocality is guaranteed by violation of Mermin and $15$-th facet inequality respectively. Interestingly, for any arbitrarily small value of $C_{GM}^{\rho(p, q)}$ there exists atleast one GHZ-symmetric state which exhibits standard nonlocality.}}
\end{figure}
\subsection{C. Relation between Genuine entanglement and Genuine nonlocality}
Till date no relationship between genuine entanglement and genuine nonlocality has been proved  even for three qubit pure quantum states. However, recently a conjecture has been reported in \cite{Ban}, which states that all three qubit genuinely entangled pure states exhibit genuine nonlocality. More recently this conjecture is proved for some important class of pure states \cite{Kau}. But no such straightforward conclusion can be drawn for any class of three qubit mixed states as there exist genuinely entangled mixed states which do not exhibit genuine nonlocality \cite{Tot}. Here we obtain the relationship between above two phenomena for three qubit GHZ-symmetric states. In this context, we have presented a subclass of states of GHZ-symmetric class of mixed tripartite states which is genuinely entangled yet fails to violate any of $185$  facet inequalities and thereby is not genuinely nonlocal (see subsection(E)).\\
However, the discussion of genuine nonlocality(Section IV) points out that out of the $185$ facet inequalities, Svetlichny inequality and the $99$-th facet inequality are the two most efficient detectors of genuine nonlocality for this class of states. In this section we have studied the variation of  violation of these two efficient bell inequalities with the amount of entanglement content $C_{GM}^{\rho(p, q)}$. Using Eq.(\ref{g14}), the maximum violation value of Svetlichny inequality(Eq.(\ref{g11})) becomes:
$S_{max} = 4\sqrt{2}(C_{GM}^{\rho(p, q)} + \frac{3}{4}-\sqrt{3}q).$
The algebraic expression clearly points out the relation between genuine nonlocality and entanglement(see Fig.6.(a)). It is already argued in Section IV, that for violation of Svetlichny inequality, the state parameters get restricted as $|p|>\frac{1}{2\sqrt{2}}$ and $q>\frac{1}{\sqrt{3}}(\frac{1}{\sqrt{2}}-\frac{1}{4})$. These restrictions, when imposed in Eq.(\ref{g14}) imply that for $C_{GM}^{\rho(p, q)}>\sqrt{2}-1$ Svetlichny inequality is violated. So any state having $C_{GM}^{\rho(p, q)}\leq \sqrt{2}-1$ cannot violate the Svetlichny inequality(see Fig.6(a)). Similarly by Eq.(\ref{g14}), the maximum violation value of $99$-th facet becomes:
\begin{equation}\label{t}
  NS_{max} = \frac{4q}{\sqrt{3}} + 2\sqrt{\frac{16 q^2}{3}+ (C_{GM}^{\rho(p, q)} + \frac{3}{4}-\sqrt{3}q)^2 } \leq 3
\end{equation}
Clearly for any arbitrary value of $q$, the amount of genuine nonlocality($NS_{max}-3$) increases monotonically with the amount of entanglement $C_{GM}^{\rho(p, q)}$. Interestingly, for any positive value of $C_{GM}^{\rho(p, q)}$, there exists a subclass which violates $99$-th facet inequality(see Fig.6(b)). To be precise, there exists a subclass of GHZ-Symmetric states which is genuinely nonlocal for any amount of $C_{GM}^{\rho(p, q)}$.
 \begin{figure}[htb]
\includegraphics[width=1.65in]{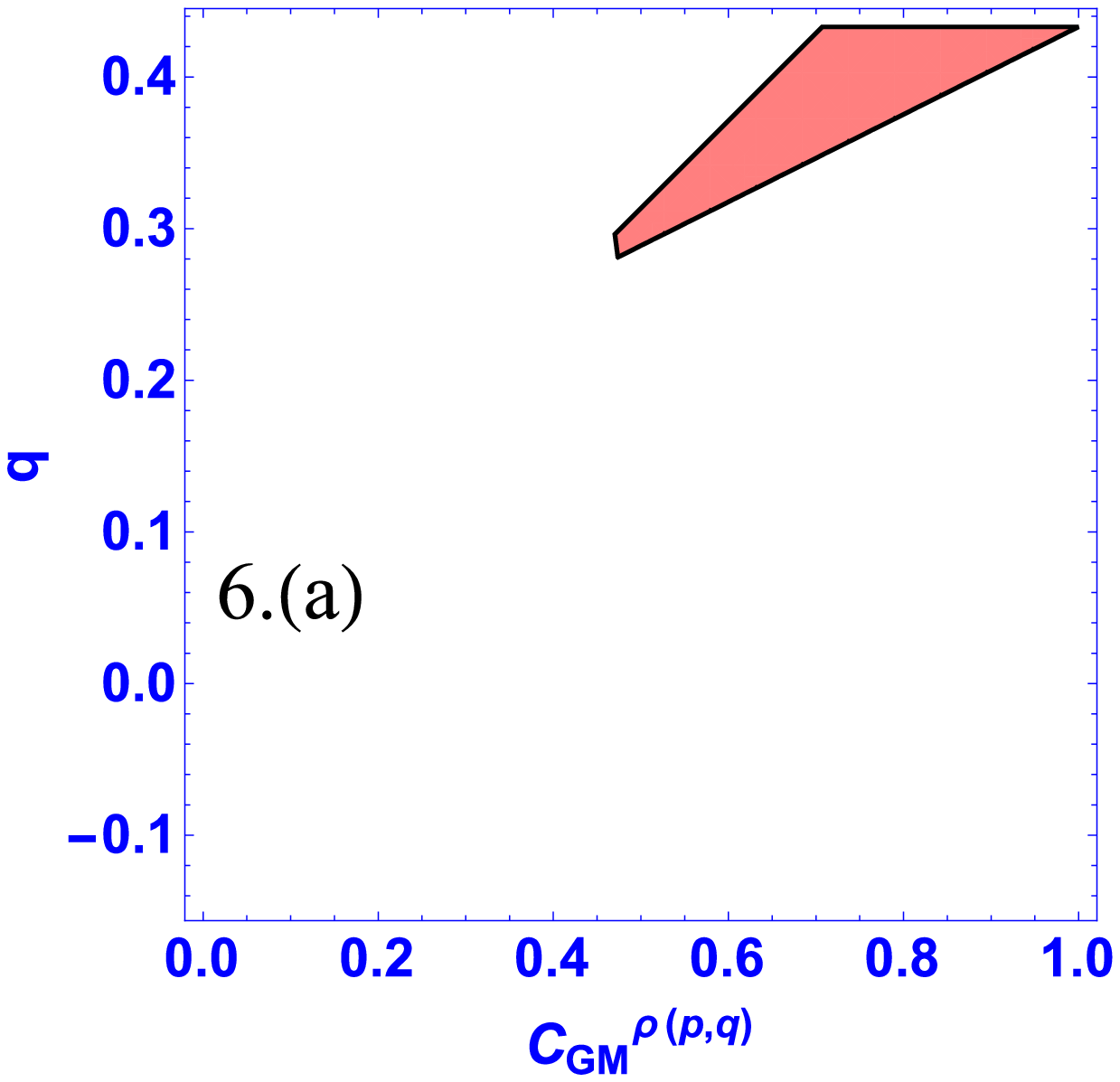}
\includegraphics[width=1.65in]{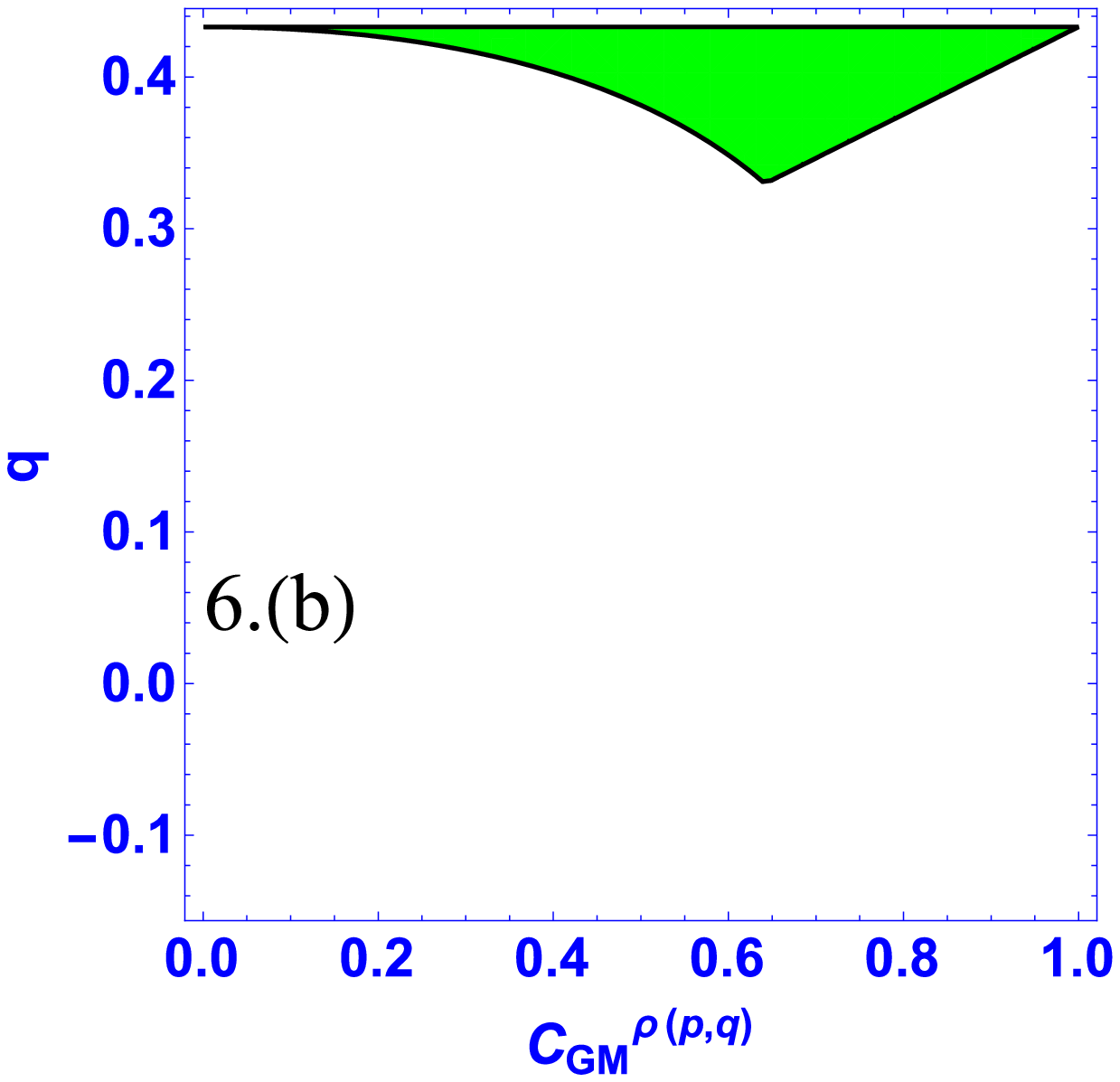}
\caption{\emph{Variation of state parameter $q$ with that of $C_{GM}^{\rho(p, q)}$ for genuinely nonlocal states are shown in these figures. The genuinely nonlocal states, as detected by Svetlichny inequality and $99$-th facet inequality are considered separately in Fig.(a) and Fig.(b) respectively. It is interesting to note that for any positive value of $C_{GM}^{\rho(p, q)}$, there exist some states whose genuine nonlocality is observed via violation of $99$-th facet. However for the states whose genuine nonlocality is guaranteed by the violation of Svetlichny inequality, no such conclusion can be made. In fact for violation of Svetlichny inequality, the range of $C_{GM}^{\rho(p, q)}$ gets restricted : $C_{GM}^{\rho(p, q)} > \sqrt{2}-1$  }}
\end{figure}
\subsection{D. Genuinely nonlocal subclass}
The genuinely nonlocal subclass is obtained for a fixed value of one of the two state parameters. Putting $q=\frac{\sqrt{3}}{4}$ in Eq.(\ref{g1}), we get
\begin{equation}\label{t1}
   \rho(p,\frac{\sqrt{3}}{4}) = (\frac{1}{2} + p)|GHZ_{+}\rangle\langle GHZ_{+}| + (\frac{1}{2} - p)|GHZ_{-}\rangle\langle GHZ_{-}|.
\end{equation}
This subclass of GHZ-Symmetric states is genuinely entangled, the amount of entanglement given by(Eq.(\ref{g14})):
\begin{equation}\label{t2}
 C_{GM}^{\rho(p,\frac{\sqrt{3}}{4})} = 2|p|
\end{equation}
The optimal region of standard nonlocality of this subclass is detected by $15$-th facet inequality:
\begin{equation}\label{t3}
 L_{max} = 4\frac{8 p^3-1}{4p^2- 1} > 4
\end{equation}
Clearly for any nonzero value of $p$, $L_{max}>4.$ The relation between entanglement($C_{GM}^{\rho(p, q)}$) and standard nonlocality is given by:
\begin{equation}\label{t4}
   4\frac{(C_{GM}^{\rho(p,\frac{\sqrt{3}}{4})})^3-1}{(C_{GM}^{\rho(p,\frac{\sqrt{3}}{4})})^2- 1} > 4.
\end{equation}
Eq.(\ref{t4}) points out that the amount of standard nonlocality($L_{max}-4$) increases monotonically with amount of entanglement($C_{GM}^{\rho(p, \frac{\sqrt{3}}{4})}$). Clearly any arbitrary amount of entanglement is sufficient for violation of the $15$-th facet inequality(see Fig.7). Similar sort of analysis can be made when we consider the stronger notion of genuine nonlocality. $99$-th facet inequality is the most efficient detector of genuine nonlocality for this subclass:
\begin{equation}\label{t5}
    NS_{max} = 1+ 2\sqrt{1+4 p^2} > 3
\end{equation}
Using Eq.(\ref{t2}), the above inequality gets modified as:
\begin{equation}\label{t6}
   1+ 2\sqrt{1+(C_{GM}^{\rho(p,\frac{\sqrt{3}}{4})})^2} >3
\end{equation}
Clearly for any arbitrary amount of $C_{GM}^{\rho(p, \frac{\sqrt{3}}{4})}$, this subclass can reveal genuine nonlocality(See Fig.7). Interestingly,  comparison between the $NS_{max}$ of this class $\rho(p, \frac{\sqrt{3}}{4})$ and that of the pure class of generalized Greenberger-Horne-Zeilinger states(GGHZ)\cite{Gho,Ghs,Kau} points out that for these two classes, genuine nonlocality varies similarly with that of their corresponding entanglement content though one of these classes is pure(GGHZ) whereas the other one is mixed\cite{Kau}.
 \begin{figure}[htb]
\includegraphics[width=2.7in]{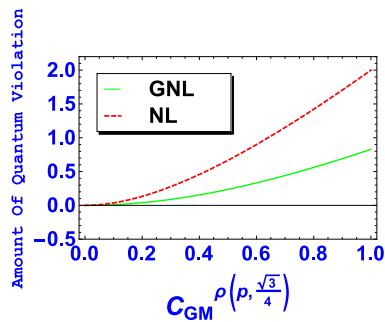}
\caption{\emph{The red dashed curve gives variation of amount of standard nonlocality($L_{max}-4$) with the amount of entanglement($C_{GM}^{\rho(p, q)}$) whereas the solid green curve represents the variation of amount of genuine nonlocality($NS_{max}-3$) with the amount of entanglement($C_{GM}^{\rho(p, q)}$). The figure shows that standard nonlocality(NL) and genuine nonlocality(GNL) both are obtained for any positive value of $C_{GM}^{\rho(p, q)}$. The curve showing variation of GNL with $C_{GM}^{\rho(p, q)}$ for this mixed subclass of GHZ-symmetric states is same as that of the curve showing variation of GNL with $C_{GM}^{\rho(p, q)}$ for  pure generalized GHZ State\cite{Kau}.}}
\end{figure}
\subsection{E. Genuinely entangled but not Genuinely nonlocal subclass}
In this subsection we present a subclass of GHZ-symmetric states which is genuinely entangled but satisfy all the $185$ facet inequalities detecting genuine nonlocality. For that we first consider Eq.(\ref{g20}) which gives the criterion of genuine entanglement: $|p| > \frac{3}{8}-\frac{\sqrt{3}}{2}q$. Now any subclass having state parameters $p$ and $q$ restricted by this criterion cannot reveal genuine nonlocality if it cannot violate neither Svetlichny inequality nor $99$-th facet inequality, i.e., if $p\leq \frac{1}{2\sqrt{2}}$ and criteria for satisfying $99$-th facet inequality: $\frac{4q}{\sqrt{3}} + 2\sqrt{\frac{16 q^2}{3}+ 4p^2 } \leq 3\,,\,q\leq\frac{1}{28}(8\sqrt{5}-5\sqrt{3}).$ Clearly these three restrictions together give a feasible region in state parameter space $(p,q)$ and that any GHZ-symmetric state having state parameter lying in this feasible region fails to reveal genuine nonlocality inspite of being genuinely entangled.
\subsection{V. CONCLUSION}
In summary, the above systematic study exploits the  nature of different notions of nonlocality thereby giving the necessary and sufficient conditions for detecting nonlocality of an entire family of high-rank mixed three-qubit states with the same symmetry as the GHZ state. Generally Mermin inequality(which is a natural generalization of  CHSH inequality) is used to detect standard nonlocality. However we have showed that this inequality is not the most efficient Bell inequality for this class of three qubit mixed states as for some restricted range of state parameters, $15$-th facet inequality gives advantage over Mermin inequality. Our findings confirm that the nonlocality conditions given by $15$-th facet inequality and Mermin inequality are the best detector of standard nonlocality for this class of states. Analogously genuine nonlocality of the class is discussed. For detection of genuine nonlocality $99$-th facet inequality and Svetlichny inequality turn out to be the most effective tools. Further comparison between these two inequalities points out that $99$-th facet inequality is even far better than Svetlichny for some restricted subclasses of this class though the latter is extensively used for detection of genuine nonlocality. Besides, our result illustrates the relationship between entanglement and nonlocality of this class of three qubit mixed states. Interestingly no biseparable state is capable of revealing standard nonlocality. This in turn points out the necessity of genuine entanglement of this class for this purpose. However for revelation of standard nonlocality existence of genuine entanglement is not sufficient. This fact becomes clear from the existence of genuine entangled local subclass of GHZ-symmetric states. It will be interesting to explore the presence of hidden nonlocality(if any) \cite{Pop,Gis,Hir} of this class of states. Also one may try to activate nonlocality of this class of states by using it in some suitable quantum network\cite{Cav,Bis}. Besides GHZ-symmetric class of states form a two dimensional affine subspace of the whole  eight dimensional space of three qubit states \cite{Goy}. So a study analyzing the relation between entanglement and nonlocality of tripartite states from other subspaces, or if possible characterization of the whole space itself can be made in future.

\section{Appendix }
In order to obtain the maximum value $M_{max}$ (Eq.(\ref{g7})) we consider the projective measurements: $A_0 = \vec{a}.\vec{\sigma_1} $ or $A_1 = \vec{\acute{a}}.\vec{\sigma_1}$ on $1^{st}$ qubit, $B_0 = \vec{b}.\vec{\sigma_2} $ or $B_1 = \vec{\acute{b}}.\vec{\sigma_2}$ on $2^{nd}$ qubit, and $C_0 = \vec{c}.\vec{\sigma_3} $ or $C_1 = \vec{\acute{c}}.\vec{\sigma_3}$ on $3^{rd}$ qubit, where $\vec{a},\vec{\acute{a}},\vec{b},\vec{\acute{b}}$ and $\vec{c},\vec{\acute{c}}$ are unit vectors and $\vec{\sigma_i}$ are the spin projection operators that can be written in terms of the Pauli matrices. Representing the unit vectors in spherical coordinates, we have, $\vec{a} = (\sin\theta a_0 \cos\phi a_0, \sin\theta a_0 \sin\phi a_0, \cos\theta a_0), ~~\vec{b} = (\sin\alpha b_0 \cos\beta b_0, \sin\alpha b_0 \sin\beta b_0, \cos\alpha b_0) $ and $\vec{c} = (\sin\zeta c_0 \cos\eta c_0, \sin\zeta c_0 \sin\eta c_0, \cos\zeta c_0) $ and similarly, we define, $\vec{\acute{a}},\vec{\acute{b}}$ and $\vec{\acute{c}}$ by replacing $0$ in the indices by $1$.  Then the value of $M$ (Eq.(\ref{g3}) of main paper) for the state $\rho(p, q)$ can be written as:
$M(\rho(p, q)) = |2p(\cos(\beta b_0 + \eta c_0 + \phi a_0)\sin(\alpha b_0) \sin(\zeta c_0) \sin(\theta a_0)- \cos(\beta b_1 + \eta c_1 + \phi a_0) \sin(\alpha b_1) \sin(\zeta c_1) \sin(\theta a_0)+ \cos(\beta b_1 + \eta c_0 + \phi a_1) \sin(\alpha b_1) \sin(\zeta c_0)\sin(\theta a_1)+$
 \begin{equation}\label{g21}
\cos(\beta b_0 + \eta c_1 + \phi a_1) \sin(\alpha b_0) \sin(\zeta c_1) \sin(\theta a_1))|
\end{equation}
To obtain the maximum value of $M$ we have to maximized  the above function $M(\rho(p, q))$ over all measurement angles.  We first find the global maximum of $M(\rho(p, q))$ with respect to $\theta a_0$ and $\theta a_1$. We begin by finding all critical points of $M(\rho(p, q))$ inside the region $R=[0 , 2\pi]\times [0 , 2\pi]$ which are namely $(\frac{\pi}{2}, -\frac{\pi}{2})$,$(-\frac{\pi}{2}, \frac{\pi}{2})$ , $(\frac{\pi}{2}, \frac{\pi}{2})$ and $(-\frac{\pi}{2}, -\frac{\pi}{2})$. The function gives maximum value with respect to $\theta a_0$ and $\theta a_1$ in all these critical points. In particular if we take $(\frac{\pi}{2}, \frac{\pi}{2})$ as the maximum critical point, then Eq.(\ref{g21}) becomes
$M(\rho(p, q)) \leq |2p(\cos(\beta b_0 + \eta c_0 + \phi a_0)\sin(\alpha b_0) \sin(\zeta c_0) - \cos(\beta b_1 + \eta c_1 + \phi a_0) \sin(\alpha b_1) \sin(\zeta c_1) + \cos(\beta b_1 + \eta c_0 + \phi a_1) \sin(\alpha b_1) \sin(\zeta c_0)+$
\begin{equation}\label{g22}
\cos(\beta b_0 + \eta c_1 + \phi a_1) \sin(\alpha b_0) \sin(\zeta c_1))|
\end{equation}
Now we carry out the same procedure over the following pair of variables $(\beta b_0, \beta b_1)$ and $(\zeta c_0, \zeta c_1)$, one by one. Similar to the previous case, critical point $(\frac{\pi}{2}, \frac{\pi}{2})$ gives the maximum value for both of these pair of variables. So the last inequality in Eq.(\ref{g22}), takes the form
\begin{equation}\label{g23}
M(\rho(p, q)) \leq |2 p G|
\end{equation}
where $G=\cos(\beta b_0 + \eta c_0 + \phi a_0) - \cos(\beta b_1 + \eta c_1 + \phi a_0)  + \cos(\beta b_1 + \eta c_0 + \phi a_1)+ \cos(\beta b_0 + \eta c_1 + \phi a_1)$.
Now the algebraic maximum value of G is equal to 4 which can be obtained by taking $\beta b_0 = 0$, $\beta b_1 = -\frac{\pi}{2}$, $\phi a_0 = 0$, $\phi a_1 = \frac{\pi}{2}$, $\eta c_0= 0$ and $\eta c_1 =  -\frac{\pi}{2}.$ Thus, $M_{max} = 8 |p|$ as obtained in Eq.(\ref{g7}). Similarly one can obtain $L_{max}$(Eq.(\ref{g10})), $S_{max}$(Eq.(\ref{g11})) and $NS_{max}$(Eq.(\ref{g13})).

\end{document}